\documentclass[10pt,conference]{IEEEtran}
\usepackage{cite}
\usepackage{amsmath,amssymb,amsfonts}
\usepackage[ruled,vlined,linesnumbered]{algorithm2e} 
\usepackage{graphicx}
\usepackage{etoolbox}
\makeatletter
\AtBeginDocument{%
  \patchcmd{\Gread@pdf}{\Gread@transgrouptrue}{\Gread@transgroupfalse}{}{}%
}
\makeatother
\usepackage{textcomp}
\usepackage{xcolor}

\usepackage[hyphens]{url}
\usepackage{fancyhdr}
\usepackage{hyperref}
\hypersetup{hidelinks}
\usepackage{booktabs}
\usepackage{multirow}
\usepackage{subcaption}
\usepackage{enumitem}
\usepackage{needspace}
\usepackage{placeins}
\usepackage{stfloats}
\usepackage{physics}
\usepackage{tikz}
\usetikzlibrary{arrows.meta,positioning}

\newcommand{\hpcayear}{2027}

\makeatletter
\newif\ifhpcahasreferences
\hpcahasreferencesfalse
\let\hpcaorig@citex\@citex
\def\@citex[#1]#2{%
  \global\hpcahasreferencestrue
  \hpcaorig@citex[{#1}]{#2}}
\let\hpcaorig@nocite\@nocite
\def\@nocite#1{%
  \global\hpcahasreferencestrue
  \hpcaorig@nocite{#1}}
\makeatother

\newcommand{\hpcasubmissionnumber}{1935}
\title{BRIDGE: Lazy-Move Compilation for Dual-Species Neutral-Atom Quantum Computers}

\def\hpcacameraready{} 

\newcommand\hpcaauthors{
Chen Huang\textsuperscript{1,2,$\dagger$},
Jingbo Wang\textsuperscript{1,$\dagger$,*},
Zhemin Zhang\textsuperscript{3},
Ming Zhong\textsuperscript{2},
Zhuo Fu\textsuperscript{4},\\
Zhiding Liang\textsuperscript{2,5,*},
Yuan Sun\textsuperscript{6,*},
Dong E. Liu\textsuperscript{7,*}%
}

\newcommand\hpcaaffiliation{%
\textsuperscript{1}Beijing Academy of Quantum Information Sciences \\
\textsuperscript{2}Department of Computer Science and Engineering, The Chinese University of Hong Kong \\
\textsuperscript{3}Open Quantum Intelligence
\textsuperscript{4}CAS Cold Atom Technology \\
\textsuperscript{5}State Key Laboratory of Quantum Information Technologies and Materials (CUHK) \\
\textsuperscript{6}Shanghai Institute of Optics and Fine Mechanics
\textsuperscript{7}Department of Physics, Tsinghua University \\
{\footnotesize $\dagger$These authors contributed equally to this work.}\
{\footnotesize *Corresponding authors.}%
}

\newcommand\hpcaemail{%
huangc@cse.cuhk.edu.hk, wangjb@baqis.ac.cn,
hisen@open-qi.com, mzhong25@cse.cuhk.edu.hk \\
fuzhuo@cascoldatom.com, zliang@cse.cuhk.edu.hk,
yuansun@live.com, dongeliu@mail.tsinghua.edu.cn
}

\author{
  \ifdefined\hpcacameraready
    \IEEEauthorblockN{\hpcaauthors{}}
      \IEEEauthorblockA{
        \hpcaaffiliation{} \\
        \hpcaemail{}
      }
  \else
    \IEEEauthorblockN{\normalsize{HPCA \hpcayear{} Submission
      \textbf{\#\hpcasubmissionnumber{}}} \\
      \IEEEauthorblockA{
        Confidential Draft \\
        Do NOT Distribute!!
      }
    }
  \fi 
}

\fancypagestyle{camerareadyfirstpage}{%
  \fancyhead{}
  
  \fancyhead[C]{
    \ifdefined\aeopen
    \parbox[][12mm][t]{13.5cm}{\hpcayear{} IEEE International Symposium on High-Performance Computer Architecture (HPCA)}    
    \else
      \ifdefined\aereviewed
      \parbox[][12mm][t]{13.5cm}{\hpcayear{} IEEE International Symposium on High-Performance Computer Architecture (HPCA)}
      \else
      \ifdefined\aereproduced
      \parbox[][12mm][t]{13.5cm}{\hpcayear{} IEEE International Symposium on High-Performance Computer Architecture (HPCA)}
      \else
      \parbox[][0mm][t]{13.5cm}{\hpcayear{} IEEE International Symposium on High-Performance Computer Architecture (HPCA)}
    \fi 
    \fi 
    \fi 
    \ifdefined\aeopen 
      \includegraphics[width=12mm,height=12mm]{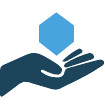}
    \fi 
    \ifdefined\aereviewed
      \includegraphics[width=12mm,height=12mm]{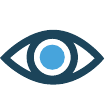}
    \fi 
    \ifdefined\aereproduced
      \includegraphics[width=12mm,height=12mm]{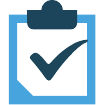}
    \fi
  }
  \fancyfoot[C]{}
}
\begin{document}

\maketitle

\ifdefined\hpcacameraready 
  \thispagestyle{camerareadyfirstpage}
  \pagestyle{empty}
\else
  \thispagestyle{plain}
  \pagestyle{plain}
\fi

\newcommand{\hpcaheight}{0mm}
\ifdefined\eaopen
\renewcommand{\hpcaheight}{12mm}
\fi

\begin{abstract}
Neutral-atom quantum computers commonly connect distant qubits by transporting atoms, incurring trap-handover errors, heating, and atom loss. We present BRIDGE (Buffer-Relay Interconnect for Data-stable Gate Execution), a hardware--software co-design that connects stationary data atoms through a fixed, second-species buffer network. Short buffer chains mediate remote controlled-Z gates and return to their initial states after each operation. BRIDGE-F combines relay routing with buffer-mediated SWAPs to relocate quantum states without transporting atoms. BRIDGE-H additionally permits data-atom transport when its full modeled cost is offset by higher survival-weighted fidelity. Across 21 circuits, BRIDGE-F exceeds ZAP and Enola in modeled survival-weighted fidelity, with geometric-mean ratios of $2.1\times$ and $38\times$ and medians of $1.3\times$ and $3.7\times$. Modeled execution latency is lower by factors of approximately $500$ and $1{,}350$, respectively, with circuit single-qubit costs excluded and relay/SWAP Hadamards included. On the same fabric, BRIDGE-F matches or exceeds SABRE-based routing on 17 circuits. Rescoring nominal plans with relay-module error equal to direct-CZ error preserves the advantage over ZAP across all 21 circuits, although recompilation can underperform on QFT. These results identify stationary buffer-mediated connectivity as a promising alternative to repeated atom transport under the stated physical assumptions.
\end{abstract}

\section{Introduction}
\label{sec:introduction}

\begin{figure}[t]
\centering
\includegraphics[width=\linewidth]{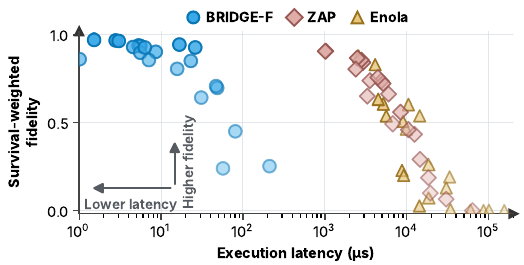}
\caption{BRIDGE-F versus reconfiguration-centric compilers (ZAP and Enola) on 21 circuits. Each marker is one method--circuit result; lighter markers indicate more native CZ gates. Markers toward the upper left have higher survival-weighted fidelity and lower execution latency. Both metrics exclude circuit single-qubit costs for all methods; BRIDGE's relay Hadamards are charged.}
\vspace{-2ex}
\label{fig:teaser}
\end{figure}

Reconfigurable connectivity is a central advantage of neutral-atom quantum computers. Optical tweezers transport atoms to bring distant qubits within range of a Rydberg gate~\cite{henrietQuantumComputingNeutral2020,saffmanQuantumInformationRydberg2010}. Each relocation, however, requires trap handovers and transport lasting orders of magnitude longer than a Rydberg gate. These operations introduce additional error, heating, and atom-loss risk~\cite{doi:10.1126/science.aah3778,bluvstein2022}. Circuits with frequently changing interaction partners incur these costs repeatedly.

We ask whether a fixed network of auxiliary atoms can provide this connectivity with less transport. In BRIDGE, buffer atoms occupy sites between data atoms and mediate remote gates through short chains. Each relay restores its buffers, while data atoms retain the algorithmic state and remain stationary unless relocation improves modeled quality. The design trades additional atoms and control resources for reduced transport, with benefits determined by the relative costs of relaying and motion.

A second atomic species separates control of the data register from that of the relay network. Species-selective addressing permits independent control~\cite{singh2022,anand2024}, while buffer-atom-mediated (BAM) gates provide a mechanism for auxiliary-mediated entanglement~\cite{sun2024}. BRIDGE interleaves Rb data atoms with Cs buffers, allowing each data atom to interact with four buffers and neighboring buffers to interact. Direct data--data interactions are excluded by the assumed interaction ranges. Buffers may become entangled with the data during a relay, provided the target gate and joint buffer recovery are achieved at completion.

Geometric reachability alone does not ensure efficient communication through this network. Each additional buffer increases modeled error and latency, while competing gates must wait for shared buffers. Dense circuits, including the quantum Fourier transform (QFT), expose these costs. Compilation must therefore coordinate placement, relay routing, and state relocation to reduce the cumulative cost of subsequent interactions.

We present \textbf{BRIDGE} (\underline{B}uffer-\underline{R}elay \underline{I}nterconnect for \underline{D}ata-stable \underline{G}ate \underline{E}xecution), a hardware--software co-design based on \emph{lazy-move} compilation. Quantum states remain in place unless relocation improves the modeled execution outcome. \textbf{BRIDGE-F} combines training-free placement, shortest-path relay routing, and buffer-mediated SWAPs without transporting atoms (\S\ref{sec:compiler}). Compatible relays execute concurrently, while SWAPs can reduce the cost of later interactions. \textbf{BRIDGE-H} extends the F plan with physical relocation of individual data atoms or small batches carried by an acousto-optic deflector (AOD). Relocation is retained only when the complete plan improves modeled survival-weighted fidelity, accounting for gate errors, idle decoherence, and atom loss. We use F and H to denote these variants throughout.

We evaluate complete architecture--compiler designs on identical native-gate workloads, using ZAP and Enola in their native configurations. Circuit single-qubit costs are excluded for every method, whereas relay and SWAP Hadamards are charged. F achieves higher modeled survival-weighted fidelity and lower execution latency on all 21 circuits (Figure~\ref{fig:teaser}). Same-fabric comparisons isolate the contributions of routing, mobility, relay construction, and placement. Hardware-cost sweeps then identify conditions under which the advantage persists or reverses. These comparisons assess the proposed co-design under explicit physical assumptions; they do not establish equal-hardware-budget superiority.

This paper makes three contributions:
\begin{itemize}[leftmargin=*,nosep]
  \item \textbf{A stationary buffer-relay architecture.} A staggered dual-species array turns relay paths into a resource scheduled by the compiler, trading auxiliary atoms for data-atom transport. An explicit relay construction requires $2L-1$ Rydberg modules and $4(L-1)$ local Hadamards for $L$ buffers, which defines the compiler's cost model.
  \item \textbf{Static-first, lazy-move compilation.} A common break-even principle evaluates state relocation through logical SWAPs or physical transport. F combines SWAP routing with training-free placement and conflict-aware scheduling. H extends the resulting static plan through incremental comparisons of moving and remaining stationary.
  \item \textbf{Performance and operating boundaries.} Same-fabric ablations assess routing, mobility, BAM, and placement choices. Sensitivity studies examine relay costs, atom loss, and pre-run vacancies, while scaling studies assess compilation up to 500 qubits.
\end{itemize}

\section{Background}
\label{sec:background}

BRIDGE combines three physical ingredients: Rydberg blockade, buffer-mediated gates, and species-selective control. Together, they define the interactions and control requirements of the proposed fabric.

\subsection{Rydberg Connectivity Constraint}
\label{subsec:blockade}

\begin{figure}[t]
\centering
\begin{subfigure}[b]{1\linewidth}
  \centering
  \includegraphics[width=\linewidth]{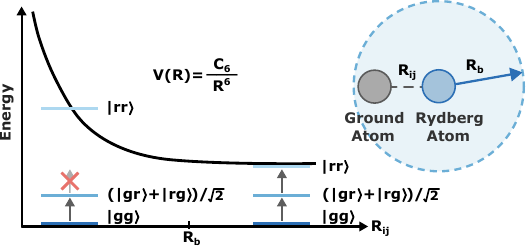}
  \caption{Rydberg blockade}
  \label{fig:blockade_mechanism}
\end{subfigure}

\vspace{0.5em}
\begin{subfigure}[b]{1\linewidth}
  \centering
  \includegraphics[width=\linewidth]{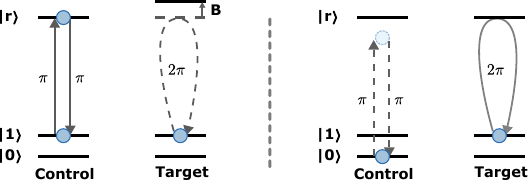}
  \caption{Native CZ gate}
  \label{fig:blockade_cz}
\end{subfigure}
\caption{%
  Rydberg blockade and native CZ gate.
  (a)~For atoms within $R_b$, the van der Waals potential $V(R)=C_6/R^6$ shifts $|rr\rangle$ off resonance, suppressing simultaneous excitation. (b)~A blockade pulse sequence converts this suppression into a state-dependent phase, yielding a native CZ for operands within $R_b$.
}
\label{fig:blockade}
\end{figure}

Neutral-atom qubits are encoded in long-lived ground-state hyperfine levels. Entangling gates transiently couple one hyperfine level to a highly excited Rydberg state. Away from resonant dipole--dipole interactions, two Rydberg-excited atoms experience the van der Waals potential
\begin{equation}
  V(R) = \frac{C_6}{R^6},
\end{equation}
where $R$ is the interatomic distance and $C_6$ is the dispersion coefficient of the chosen Rydberg level. Exciting one atom shifts the doubly excited state $|rr\rangle$ off resonance. When this shift exceeds the drive linewidth, simultaneous excitation is suppressed by Rydberg blockade (Figure~\ref{fig:blockade_mechanism}). The blockade radius $R_b$ denotes the distance below which this suppression is reliable.

Within $R_b$, state-dependent blockade supports a native controlled-Z (CZ) gate (Figure~\ref{fig:blockade_cz}). The control atom interrupts the target's $2\pi$ Rabi cycle, producing a conditional phase. Subsequent protocols, including off-resonant modulated driving (ORMD) with smooth pulses, improve CZ fidelity~\cite{sun2020controlled,fu2022high,evered2026high}. These protocols nevertheless require a strong interaction between the gate operands.

Because $V(R) \propto R^{-6}$ decays steeply, only nearby pairs can be gated reliably, and every long-range gate requires either a move or an intermediate relay. 

\subsection{Buffer-Atom-Mediated (BAM) Gates}
\label{subsec:bam}

\begin{figure}[t]
    \centering
    \includegraphics[width=\linewidth]{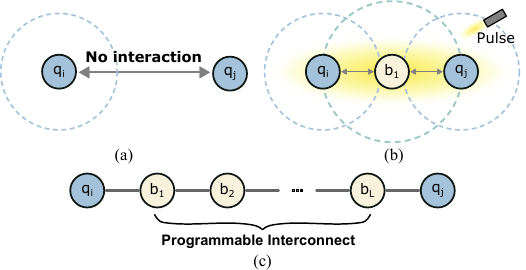}
    \caption{Buffer-mediated connectivity. (a)~Data endpoints lie outside direct interaction range. (b)~One buffer mediates the gate. (c)~A chain of $L$ buffers implements a remote CZ with $2L-1$ Rydberg modules and restores the buffers.}
    \label{fig:bam_mechanism}
\end{figure}

A BAM gate~\cite{sun2024} entangles two data atoms through an auxiliary buffer when the data atoms cannot interact directly. Each data atom lies within the buffer's blockade radius, while their separation exceeds the direct data--data blockade radius (Figure~\ref{fig:bam_mechanism}a,b). With the buffer initialized in $\ket{1}$, ORMD pulses apply a CZ phase to the data atoms and restore the buffer. A single buffer thus approximately doubles the reach of a native CZ without atom transport. Published sample waveforms yield simulated gate errors below $10^{-4}$, but experimental demonstration remains outstanding. For longer distances, the same work outlines two unsimulated approaches: measuring entangled buffers or coherently driving a chain that returns to its initial state. BRIDGE follows the coherent approach through the explicit $L$-buffer construction in \S\ref{subsec:relay}.

\subsection{Dual-Species Substrate}
\label{subsec:substrate}

BRIDGE uses $^{87}$Rb data atoms and $^{133}$Cs buffers for species-selective control. Their distinct optical and Rydberg transitions permit separate addressing of the data and buffer registers. Independent loading, imaging, and Rydberg control have been demonstrated in dual-species Rb/Cs arrays~\cite{singh2022,anand2024}. Interspecies F\"orster resonances can strengthen Rb--Cs coupling~\cite{ireland2024,anand2024}. However, Rb--Cs, Cs--Cs, and Rb--Rb interactions depend on a shared choice of Rydberg states and must be selected jointly. The proposed design uses this selectivity to place buffers near data atoms, operate the buffer network while data atoms idle, and transport Rb independently.

The buffer-relay fabric requires the interaction ranges to satisfy
\begin{equation}
  R_{b,\mathrm{Rb\text{-}Cs}} > d_{\mathrm{DB}}, \quad
  R_{b,\mathrm{Cs\text{-}Cs}} > d_{\mathrm{BB}}, \quad
  R_{b,\mathrm{Rb\text{-}Rb}} < d_{\mathrm{DD}},
  \label{eq:channel}
\end{equation}
where $d_{\mathrm{DB}}$, $d_{\mathrm{BB}}$, and $d_{\mathrm{DD}}$ are the data--buffer, buffer--buffer, and data--data spacings. Each data atom must interact with its nearest buffers, and neighboring buffers must support chain formation. Direct data--data interactions are excluded, so remote gates use the buffer network.

Selective driving of Cs atoms along each relay path requires site-resolved addressing~\cite{graham2019}. H additionally requires Rb-selective AOD transport that leaves the Cs lattice undisturbed. The scheduler excludes concurrent relays that share atoms or have crossing paths.

\section{Architecture and Execution Model}
\label{sec:architecture}

BRIDGE uses a fixed array geometry with circuit-dependent atom occupancy. We define the relay construction, array layout, and execution model used to compare schedules. The model combines atom survival with conditional circuit fidelity to assess the cost of relaying and relocation.

\subsection{Multi-Buffer Relay}
\label{subsec:relay}

A relay path $P=(q_s,b_1,\ldots,b_L,q_t)$ connects data endpoints $q_s$ and $q_t$ through $L$ buffers. The operation must apply a CZ to the endpoints and restore all buffers to the prescribed state $|\beta_P\rangle_B$. Up to a global phase, this requirement is expressed as
\begin{equation}
 U_P\bigl(|\psi\rangle_{st}|\beta_P\rangle_B\bigr)
 = \mathrm{CZ}_{st}|\psi\rangle_{st}|\beta_P\rangle_B
 \label{eq:relay_interface}
\end{equation}
for any two-qubit input $|\psi\rangle_{st}$. For the construction below, $|\beta_P\rangle_B=|0\rangle^{\otimes(L-1)}\otimes|1\rangle$, with the final buffer mediating BAM and the others coherently encoding the source value. Intermediate data--buffer entanglement is permitted, provided the endpoint phase and joint buffer recovery satisfy Eq.~\eqref{eq:relay_interface}.

The construction comprises coherent propagation, phase application, and uncomputation. First, $L-1$ sequential CNOTs encode the source value $x$ in buffer $b_{L-1}$. A BAM gate then couples $b_{L-1}$ and $q_t$ through $b_L$, imprinting the phase $(-1)^{xy}$ for target value $y$. Reversing the CNOT sequence restores the buffers while preserving this phase. For $L=1$, a single BAM gate directly couples $q_s$ and $q_t$ through the buffer. Each operation acts on neighboring atoms, so no individual Rydberg interaction spans the chain. Buffer recovery follows from uncomputation rather than measurement or reset. Physical errors are not reversed and are included in the relay cost model.

We assessed the $L=2$ construction by simulating four three-level atoms in the full 81-dimensional Hilbert space. The scalar interaction model retained all six pair interactions and leakage into multiply excited states. It yielded a joint target-state fidelity of $0.99999998$ for the endpoint CZ and buffer recovery. With a $100~\mu\mathrm{s}$ Rydberg lifetime and emission treated as failure, fidelity was $0.9966$, corresponding to $0.9989$ per module. Three Rydberg modules required $0.75~\mu\mathrm{s}$, while four $1~\mu\mathrm{s}$ Hadamards increased the total duration to $4.75~\mu\mathrm{s}$. The simulation did not resolve species-specific Rb and Cs interactions.

We define a \emph{module} as one Rydberg-mediated CZ within a CNOT, or one BAM gate. The construction requires $(L-1)+1+(L-1)=2L-1$ modules acting at fixed nearest-neighbor spacings (Figure~\ref{fig:bam_mechanism}c). Each of the $2(L-1)$ CNOTs requires two Hadamards, giving $4(L-1)$ local gates and none for $L=1$. Assuming serial operations and independent errors, the compiler assigns the costs
\begin{align}
  \widehat F_{\mathrm{BAM}\mid\mathrm{surv}}(P)
      &= F_{\mathrm{relay}}^{2L-1}\,F_{\mathrm{H}}^{4(L-1)},\label{eq:fbam}\\
  \widehat T_{\mathrm{BAM}}(P)
      &= (2L-1)\tau_p+4(L-1)\tau_{\mathrm{H}},\label{eq:tbam}
\end{align}
where $F_{\mathrm{relay}}$ and $\tau_p$ denote the survival-conditioned quality and duration of one relay module, termed a ``relay module'' in the plots. The corresponding Hadamard parameters are $F_{\mathrm{H}}$ and $\tau_{\mathrm{H}}$. Errors compound as modules accumulate, without reducing the assumed interaction strength within each module. The model excludes coherent errors, buffer dephasing during propagation, and crosstalk. Whole-chain pulse optimization could improve on the sequential construction modeled here.

The nominal relay parameters are modeling assumptions, rather than experimentally calibrated values. The assumed $F_{\mathrm{relay}}=0.997$ is below the simulated $L=2$ per-module value of $0.9989$ but above the baseline direct-CZ quality of $0.995$. The module duration $\tau_p=0.25~\mu\mathrm{s}$ covers its Rydberg-mediated component. Relay Hadamards are charged separately at $F_{\mathrm{H}}=0.9997$~\cite{evered2023} and $\tau_{\mathrm{H}}=1~\mu\mathrm{s}$, and dominate the duration of longer relays.

\subsection{Array Layout}
\label{sec:array_layout}

\begin{figure}[t]
\centering
\includegraphics[width=0.9\linewidth]{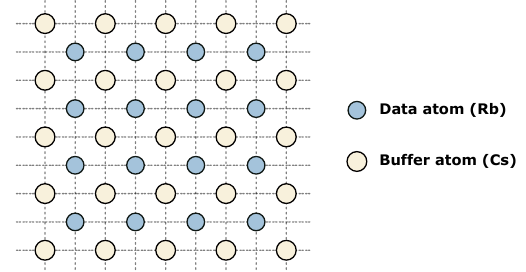}
\caption{Candidate sites of the staggered Rb--Cs array. Data and buffer site pitches are $8.0~\mu\mathrm{m}$, with $5.66~\mu\mathrm{m}$ between neighboring data and buffer sites. Only atoms required by the mapping and relay paths are retained.}
\label{fig:architecture}
\end{figure}

Figure~\ref{fig:architecture} shows the candidate sites in BRIDGE's staggered array. Cs buffer sites form a square lattice with pitch $d_{\mathrm{BB}}=8.0~\mu\mathrm{m}$, while Rb data sites occupy the plaquette centers. Thus, $d_{\mathrm{DD}}=8.0~\mu\mathrm{m}$ and $d_{\mathrm{DB}}=d_{\mathrm{BB}}/\sqrt{2}=5.66~\mu\mathrm{m}$. Under the assumed blockade radii (Table~\ref{tab:hw_params}), these spacings satisfy Eq.~\eqref{eq:channel}. Data--buffer and buffer--buffer interactions are available, while direct data--data interactions are excluded. Four equidistant buffer neighbors per data site provide a regular routing geometry, and the buffer pitch supports short paths between nearby data atoms.

Circuit placement determines which data sites are occupied in the array. Routing retains only buffers required by at least one relay, including relays following planned moves. All remaining buffer sites are left empty during modeled execution. Every scheduled relay must follow a legal path through the occupied sites.

\subsection{Atom Loss and Survival Probability}
\label{sec:survival}

The survival probability $P_{\mathrm{surv}}$ describes completion without loss of any required atom. Data-atom loss terminates the shot because replacement cannot recover an unknown quantum state. Buffer loss during a coherent relay likewise terminates the shot. A vacancy detected before execution can instead be excluded from routing if a legal alternative path remains. Refilling between shots~\cite{chiu2025} may use transport while preserving F's motion-free execution.

We estimate survival for a fixed schedule from two independent loss sources:
\begin{equation}
  \widehat P_{\mathrm{surv}}
  =\exp(-\sum_a t_a/\tau_a)\prod_i(1-p_i),
  \label{eq:survival_model}
\end{equation}
Here, $t_a$ is the required trapping time of atom $a$, including transport, and $\tau_a$ is its background lifetime. The factor $p_i$ is the additional loss probability of operation $i$, counted once for each gate, handover, or transport. All data atoms and retained buffers are exposed throughout execution, and any retained-atom loss is treated as failure. This conservative assumption excludes possible rerouting around idle buffers following loss detection.

\subsection{Conditional Fidelity and Joint Quality}
\label{sec:fidelity_model}

Even without atom loss, gates, dephasing, and leakage can degrade the output state. We define \emph{conditional circuit fidelity} $F_{\mathrm{cond}}$ relative to the ideal output, conditioned on survival. To compare schedules without state simulation, we approximate this quantity by multiplying operation-level quality factors:
\begin{equation}
  \widehat F_{\mathrm{cond}}
  =F_{\mathrm{1Q}}F_{\mathrm{relay}}^{\mathrm{circ}}F_{\mathrm{move}}F_{\mathrm{idle}},
  \label{eq:conditional_model}
\end{equation}
where the factors account for single-qubit gates, relays, relocation, and idle periods. Longer relays reduce $F_{\mathrm{relay}}^{\mathrm{circ}}$, whereas relocation introduces error but can shorten subsequent relays. All factors are conditioned on survival, avoiding double counting of loss in $\widehat P_{\mathrm{surv}}$. Relay quality includes buffer-return error; the product approximation neglects error correlations and propagation.

Relocation can improve conditional fidelity by shortening subsequent relays while reducing atom survival. We capture this trade-off using the \emph{survival-weighted fidelity}
\begin{equation}
  \widehat F_{\mathrm{sw}}=\widehat P_{\mathrm{surv}}\widehat F_{\mathrm{cond}},
  \label{eq:survival_weighted_fidelity}
\end{equation}
which combines survival with conditional fidelity and generally differs from the probability of a correct algorithmic output.

\section{The BRIDGE Compiler}
\label{sec:compiler}

\begin{figure}[t]
\centering
\includegraphics[width=\linewidth]{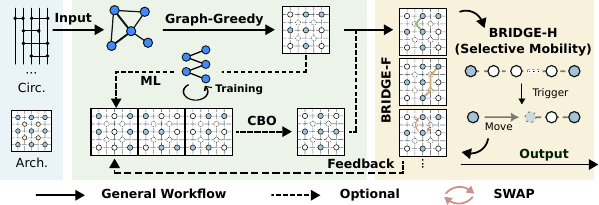}
\caption{BRIDGE compiler workflow. Graph-Greedy placement and BRIDGE-F routing produce a static plan using relays and, where beneficial, buffer-mediated SWAPs. Optional CBO search refines placement by evaluating complete static plans, with or without a learned initial candidate (dashed arrows). BRIDGE-H evaluates physical relocation against the static plan and accepts only improvements in modeled survival-weighted fidelity.}
\vspace{-2ex}
\label{fig:workflow}
\end{figure}

BRIDGE proceeds through three compilation stages (Figure~\ref{fig:workflow}). Graph-Greedy placement followed by static routing and scheduling produces the complete F plan. Selective mobility then produces H by extending the relocation decision from logical SWAPs to physical transport. H accepts a move only when the resulting complete plan strictly improves modeled survival-weighted fidelity.

\subsection{Initial Qubit Placement}
\label{sec:placement}

Placement assigns each logical qubit to a legal data site. Graph-Greedy selects the unplaced qubit with the most interactions with already placed qubits. It assigns that qubit to the free site minimizing the corresponding weighted relay cost. Ties follow weighted degree and qubit index, then distance to the array center and site coordinates. A local hill-climb next places qubits near strongly interacting partners whenever this reduces the interaction-weighted relay cost. Two forward--backward SWAP-routing passes further refine the layout, each initialized from the preceding pass's final mapping, as in SABRE~\cite{sabre2019}. A schedule-free relay-cost estimate ranks the candidates, and full evaluation compares the original Graph-Greedy layout with the two highest-ranked alternatives. For each layout $\pi$, F routing and scheduling generate a static plan $S_\pi$, scored as
\begin{equation}
\begin{aligned}
\mathcal L_{\mathrm{init}}(\pi)&=-\log\widehat F_{\mathrm{sw}}(S_\pi)\\
&=-\log\widehat P_{\mathrm{surv}}(S_\pi)
-\log\widehat F_{\mathrm{cond}}(S_\pi).
\end{aligned}
\label{eq:placement_obj}
\end{equation}
This objective includes modeled gate errors, operation-induced loss, buffer use, idle decoherence, and background exposure. Infeasible plans receive infinite cost, and placement does not anticipate subsequent physical relocation.

Optional consensus-based optimization (CBO) tests whether placement search justifies its additional compilation cost. A population of layouts evolves through elite majority votes, random proposals, and mutations guided by scheduling-blockage traces. Each candidate is scored using Eq.~\eqref{eq:placement_obj}, with latency breaking quality ties. Greedy+CBO includes the Graph-Greedy layout; Greedy+ML+CBO additionally replaces one random candidate with a learned layout. Both variants use identical population sizes, search budgets, and random seeds. The selected layout provides a common starting point for F and H.

The learned initializer uses ridge regression on qubit and site features and their cross terms. Qubit features comprise weighted degree and neighbor count; site features comprise normalized coordinates and distance from the center. Training uses random legal layouts labeled by local relay cost, followed by greedy decoding under spacing constraints. Because these labels omit scheduling delays and global exposure, complete-plan evaluation determines the final selection. The model requires retraining when the circuit, geometry, or costs change; default Graph-Greedy placement requires no training.

\begin{figure*}[t]
  \centering
  \includegraphics[width=0.86\textwidth]{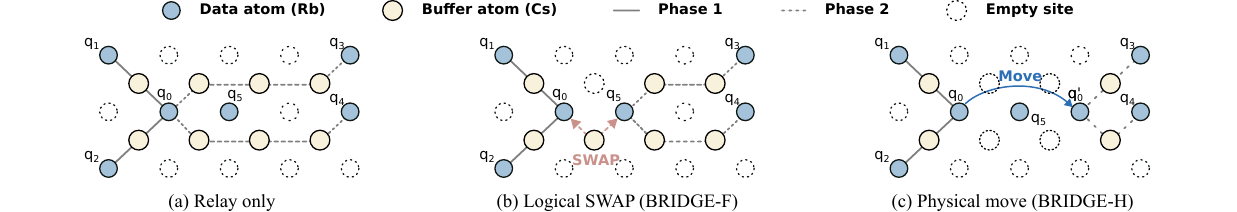}
  \caption{Three responses to changing interaction partners. Qubit $q_0$ interacts with $q_1$ and $q_2$ in Phase~1, then with distant partners $q_3$ and $q_4$ in Phase~2. (a)~Relay-only routing preserves the state mapping, requiring three buffers per Phase-2 gate. (b)~BRIDGE-F swaps the states of $q_0$ and $q_5$ through their shared buffer, using three one-buffer relays without atom transport. Subsequent gates use two-buffer relays from the atom initially carrying $q_5$. (c)~Physical relocation places the $q_0$ atom near its Phase-2 partners but incurs handover, transport-loss, and timing costs. H evaluates such moves from the F plan and accepts only improvements in modeled survival-weighted fidelity.}
  \vspace{-2ex}
  \label{fig:phase_shift}
\end{figure*}

\subsection{BRIDGE-F: Fully Fixed Routing and Scheduling}
\label{sec:static_evaluator}

With data positions fixed, each CZ uses a path minimizing the number of intermediate buffers $L$, with costs given by Eqs.~\eqref{eq:fbam}--\eqref{eq:tbam}. The router retains a bounded set of equally short alternatives. Longer detours are excluded even when they could reduce waiting, consistently across F, placement evaluation, and H.

\textbf{Relay or SWAP.} Relaying preserves the state mapping but repeatedly incurs the cost of distant interactions. A buffer-mediated SWAP can instead bring a state closer to its subsequent partners without transporting atoms (Figure~\ref{fig:phase_shift}b). For data atoms sharing a buffer, one SWAP requires three one-buffer relays and six local Hadamards, using CNOT decomposition H$\cdot$CZ$\cdot$H. These Hadamards incur the same costs as relay Hadamards. Before a CZ requiring at least two buffers, F considers swapping either operand with an adjacent data atom. Candidates are ranked by savings over the next 20 CZ gates, with each successive gate weighted by a further factor of $0.9$. The best candidate is accepted only when these savings exceed the SWAP cost. This step repeats at most eight times per gate, stopping when no candidate is beneficial. The scheduler then evaluates the modified gate stream, retaining it only if the complete plan improves $\widehat F_{\mathrm{sw}}$ over relay-only routing.

The event-driven scheduler jointly selects relay paths whenever gates become ready or operations finish. It first maximizes the number of launched gates, then prioritizes the longest remaining critical path, with identifiers breaking remaining ties. Running relays reserve their endpoints and buffers, and concurrent paths cannot cross. If A--B has two shortest paths and C--D has one, the scheduler assigns A--B the path that leaves C--D available. Gates wait when no compatible combination exists under these constraints. This heuristic does not guarantee an optimal makespan; its blocking traces guide subsequent CBO and H proposals.

\subsection{BRIDGE-H: Selective Mobility via Incremental Evaluation}
\label{sec:hybrid_evaluator}

Changing interaction partners can make an initially favorable layout inefficient later in a circuit. In Figure~\ref{fig:phase_shift}, data-atom $q_0$ interacts first with $q_1$ and $q_2$, then with distant partners $q_3$ and $q_4$. F can swap the state closer to these partners; H also evaluates physical relocation. At operation-completion events, H ranks data atoms with long or blocked relays and vacant sites near their upcoming partners. Local relay savings rank candidates, while the complete execution model determines acceptance. Before an accepted move, H stops new gate launches and allows running gates to finish. It then relocates the atom to a legal vacant site and resumes F scheduling.

A relocation over distance $d$ takes $T_{\mathrm{move}}=2t_{\mathrm{ho}}+2\sqrt{d/a}$, assuming symmetric acceleration and deceleration. This includes two handovers between spatial-light-modulator (SLM) traps and moving AOD traps. The model charges a handover quality factor $F_{\mathrm{ho}}^2$, moving-atom dephasing, and transport loss $p_{\mathrm{move}}$ once per relocation. All other atoms remain idle during transport, after in-flight gates have completed. Rb-selective AOD batches may move several data atoms simultaneously while preserving row and column order. Batch motion must also avoid activating AOD traps at stationary data-atom sites.

A local estimate illustrates the balance between relay savings and relocation cost. Removing one buffer saves two modules and four Hadamards, improving the quality factor by $F_{\mathrm{relay}}^{-2}F_{\mathrm{H}}^{-4}$. Relocation incurs the factor $F_{\mathrm{ho}}^2(1-p_{\mathrm{move}})$, giving a nominal break-even saving of approximately $0.4$ buffers when timing costs are omitted. Spectator waiting during the $\ge30~\mu\mathrm{s}$ relocation raises this requirement. Although a move incurs less local error than a SWAP (approximately $0.3\%$ versus $1.1\%$), it stalls other atoms and risks loss. These costs restrict mobility gains when relays already use one buffer or SWAP routing has shortened distant interactions (\S\ref{sec:eval_placement}).

H compares each candidate plan $S_m$ with the better of the current plan and a rebuilt plan without relocation. This reference plan $S_s$ is shared across candidates at the same event. For $C(S)=-\log\widehat F_{\mathrm{sw}}(S)$, the difference $\Delta C_m=C(S_m)-C(S_s)$ separates into gate, movement, buffer-reuse, idle, and background terms. Contributions from unchanged gates and the completed prefix cancel between plans. H updates affected relay costs and propagates scheduling changes through dependencies and shared buffers. When the revised schedule matches an existing checkpoint, the remaining schedule is reused with a time shift. Incrementally updated makespan, busy times, and retained-buffer count determine global costs, allowing a constant-time final acceptance test after rescheduling.

H accepts only feasible candidates satisfying $\Delta C_m<0$ and selects the largest quality improvement. Latency and then identifier resolve ties before compilation continues from the updated schedule. A latency-only improvement cannot authorize relocation, and atoms remain stationary when no candidate improves quality.

\section{Experimental Methodology}
\label{sec:methodology}

\subsection{Benchmarks, Baselines, and Resources}

We evaluate 21 circuits: 16 from QASMBench~\cite{li2023} and five from VeriQBench~\cite{chen2022}. All methods receive the same native-gate workload; Table~\ref{tab:resources} lists its qubit and CZ counts.

We compare three architecture--compiler co-designs, each in its native configuration. ZAP~\cite{huang2026} uses a zoned architecture and Enola~\cite{tan2025} an all-in-one architecture; both connect qubits by moving atoms. BRIDGE uses a fixed buffer-relay fabric with optional data-atom relocation. The comparison therefore evaluates complete designs rather than compilers alone; \S\ref{sec:eval_placement} isolates BRIDGE's compiler choices on the same fabric.

For each circuit, we count the $N$ data atoms loaded by all methods and the $B$ Cs buffers retained by BRIDGE's plan. Trap counts include distinct sites where atoms rest during execution, including initially empty entanglement or partner sites used later. Spatial footprint is the bounding-box area enclosing all atom positions throughout execution, including parking and entanglement locations. We compare footprints per circuit relative to BRIDGE-F and aggregate the ratios geometrically. The quality model includes the additional buffer population. BRIDGE also requires a second atomic species, a separate set of Rydberg lasers, and site-selective control of Cs paths.

\subsection{Hardware Parameters and Evaluation Metrics}

Table~\ref{tab:hw_params} lists the nominal parameters used in the evaluation. Gate and coherence parameters follow neutral-atom experiments~\cite{levine2019,evered2023,bluvstein2022}; relay parameters are assumptions assessed by sensitivity analysis (\S\ref{subsec:relay}). Gate and handover qualities are conditioned on survival, with atom loss charged separately. Cooled dual-species arrays motivate the nominal $20$~s background lifetime~\cite{singh2022}, varied in \S\ref{sec:eval_loss}. ZAP and Enola share BRIDGE's handover, coherence, lifetime, and transport-loss parameters. Their direct CZ has assumed quality $0.995$ and duration $0.36~\mu\mathrm{s}$.

\begin{table}[t]
\centering
\caption{Nominal evaluation parameters. Circuit single-qubit costs are excluded; relay and SWAP Hadamards are charged.}
\label{tab:hw_params}
\footnotesize
\setlength{\tabcolsep}{15pt}
\begin{tabular}{@{}ll@{}}
\toprule
Parameter & Value \\
\midrule
Data / buffer species & $^{87}$Rb / $^{133}$Cs \\
Data--buffer / same-species spacing & $5.66$ / $8.0~\mu\mathrm{m}$ \\
Rb--Rb / Rb--Cs / Cs--Cs model radii & $5$ / $10$ / $9~\mu\mathrm{m}$ \\
Circuit single-qubit gate quality / duration & $1$ / $0$ \\
Baseline direct CZ quality / duration & $0.995$ / $0.36~\mu\mathrm{s}$ \\
Relay module quality / duration & $0.997$ / $0.25~\mu\mathrm{s}$ \\
Relay/SWAP Hadamard quality / duration & $0.9997$ / $1~\mu\mathrm{s}$ \\
Handover quality (survival-conditioned) / time & $0.999$ / $15~\mu\mathrm{s}$  \\
Transport acceleration bound & $2.75~\mu\mathrm{m}/\mu\mathrm{s}^{2}$ \\
Data coherence time $T_2$ & $1.5~\mathrm{s}$ \\
\midrule
Additional gate-induced atom loss & $0$ \\
Rb / Cs background lifetime $\tau_a$ & $20$ / $20~\mathrm{s}$ \\
Additional loss per relocation $p_{\mathrm{move}}$ & $10^{-3}$ \\
\bottomrule
\end{tabular}
\end{table}

Quality denotes estimated survival-weighted fidelity, $F_{\mathrm{sw}}=P_{\mathrm{surv}}F_{\mathrm{cond}}$. We also report its two factors, modeled execution latency $T_{\mathrm{exec}}$, realized movement count, and compilation time. Execution latency includes relays, direct CZ gates, relay/SWAP Hadamards, transport, handovers, and waiting. It excludes circuit single-qubit costs, preparation, readout, and between-shot recovery, and therefore does not represent a complete shot cycle.

\subsection{Experimental Protocol}

\textbf{Single-qubit costs.} All methods retain the same circuit single-qubit gates and dependencies, but assign zero error, duration, and loss to these operations. This convention also applies during placement and mobility evaluation; relay and SWAP Hadamards remain fully charged (\S\ref{subsec:relay}). A common evaluator scores every exported timeline, with all data atoms exposed throughout execution. Transport loss is charged once per pickup-to-dropoff episode, separately from handover error.

\textbf{Runs and statistics.} Graph-Greedy F/H, ZAP, and Enola are deterministic and are run once per condition. Only CBO variants use random seeds, fixed at $\{7,11,23\}$. Cross-circuit summaries use geometric means over the common completed set, with medians also reported. Failures are recorded separately from completed runs with low modeled quality.

\textbf{Search limits.} CBO uses 20 population members, up to 60 iterations, and at most 256 distinct layout evaluations. F retains up to 16 equally short paths per endpoint pair. At each scheduling event, it explores up to 4096 joint-search nodes per conflict group. SWAP insertion is limited to eight operations before each gate. At each mobility event, H ranks four atoms with eight candidate sites each and fully evaluates up to eight single moves. It evaluates up to eight additional candidates for extending the best move into a batch.

\textbf{Compilation time.} $T_{\mathrm{comp}}$ measures wall-clock time for model loading, any ML training and inference, placement, routing, scheduling, and H search. It excludes front-end preprocessing, final validation, simulation, and result export. Runs use one process with single-threaded numerical libraries on a 2.30-GHz Intel Xeon cloud-server core. At most two experiments run concurrently on separate cores; scalability runs use a separate 2.80-GHz Xeon instance. Only the placement study uses ML, with training included in its reported time (\S\ref{sec:eval_placement}). All other experiments use the training-free Graph-Greedy placement procedure described above.

\section{Evaluation Results}
\label{sec:evaluation}

We first compare nominal performance (\S\ref{sec:eval_overall}), then test sensitivity to relay costs, atom loss, and transport (\S\ref{sec:eval_sensitivity}, \S\ref{sec:eval_loss}). We next isolate component contributions on the same fabric (\S\ref{sec:eval_placement}) and assess compilation scalability (\S\ref{sec:eval_scalability}). Throughout this section, quality denotes the modeled $F_{\mathrm{sw}}$.

\begin{table*}[t]
\centering
\caption{Per-circuit resources and nominal results. $N$: Rb data atoms (common to all methods); $B$: Cs buffers retained by BRIDGE (F/H where they differ); CZ: native CZ count; SW: logical SWAPs inserted by F; $\bar L$, $L_{\max}$: mean and maximum buffers per relay in the F plan. $F_{\mathrm{sw}}$ and $T_{\mathrm{exec}}$ are model estimates excluding circuit single-qubit costs; F's latency is given in $\mu$s and the baselines' in ms. Moves counts atom transport episodes; F has none. Traps: sites in which an atom rests (BRIDGE: $N+B$). Area: bounding box of all atom positions, relative to F. H differs from F only on \texttt{multiplier\_n15} (\S\ref{sec:eval_placement}).}
\label{tab:resources}
\footnotesize
\setlength{\tabcolsep}{2.75pt}
\begin{tabular}{@{}lrrrrrr|rrr|rrr|rr|rr|rr@{}}
\toprule
 & & & & & & & \multicolumn{3}{c|}{$F_{\mathrm{sw}}$} & \multicolumn{3}{c|}{$T_{\mathrm{exec}}$} & \multicolumn{2}{c|}{Moves} & \multicolumn{2}{c|}{Traps} & \multicolumn{2}{c}{Area ($\times$F)} \\
Circuit & $N$ & $B$ & CZ & SW & $\bar L$ & $L_{\max}$ & F & ZAP & Enola & F ($\mu$s) & ZAP (ms) & Enola (ms) & ZAP & Enola & ZAP & Enola & ZAP & Enola \\
\midrule
\texttt{adder\_n4} & 4 & 3 & 10 & 0 & 1.00 & 1 & 0.970 & 0.907 & 0.874 & 1.5 & 1.0 & 2.6 & 15 & 19 & 12 & 7 & 3.67 & 3.3 \\
\texttt{adder\_n64} & 64 & 53 & 455 & 5 & 1.00 & 1 & 0.241 & $3.8\mathrm{e}{-4}$ & $1.3\mathrm{e}{-17}$ & 56.2 & 62.5 & 153.1 & 668 & 909 & 130 & 183 & 1.61 & 8.3 \\
\texttt{bv\_n14} & 14 & 8 & 13 & 0 & 1.38 & 2 & 0.927 & 0.842 & 0.537 & 25.8 & 2.9 & 5.6 & 26 & 25 & 19 & 28 & 0.77 & 5.1 \\
\texttt{cat\_n22} & 22 & 18 & 21 & 0 & 1.00 & 1 & 0.939 & 0.735 & 0.229 & 5.2 & 4.8 & 8.8 & 42 & 41 & 26 & 43 & 0.58 & 7.1 \\
\texttt{cat\_n35} & 35 & 29 & 34 & 0 & 1.00 & 1 & 0.903 & 0.559 & 0.027 & 8.5 & 8.3 & 14.3 & 68 & 67 & 39 & 69 & 0.71 & 8.6 \\
\texttt{cc\_n12} & 12 & 7 & 12 & 0 & 1.25 & 2 & 0.944 & 0.856 & 0.604 & 16.5 & 2.7 & 5.2 & 24 & 23 & 18 & 23 & 1.03 & 5.2 \\
\texttt{eff\_n12} & 12 & 9 & 11 & 0 & 1.00 & 1 & 0.967 & 0.867 & 0.630 & 2.8 & 2.5 & 4.5 & 22 & 21 & 18 & 23 & 1.03 & 5.2 \\
\texttt{ghz\_n23} & 23 & 19 & 22 & 0 & 1.00 & 1 & 0.936 & 0.722 & 0.201 & 5.5 & 5.1 & 9.2 & 44 & 43 & 27 & 45 & 0.74 & 6.0 \\
\texttt{grover\_n15} & 15 & 13 & 144 & 1 & 1.00 & 1 & 0.642 & 0.188 & $2.2\mathrm{e}{-3}$ & 30.5 & 18.1 & 59.3 & 191 & 287 & 30 & 42 & 1.21 & 9.2 \\
\texttt{grover\_n9} & 9 & 7 & 72 & 0 & 1.00 & 1 & 0.805 & 0.435 & 0.130 & 15.5 & 12.3 & 30.0 & 125 & 143 & 21 & 26 & 2.10 & 7.7 \\
\texttt{ising\_n26} & 26 & 22 & 50 & 0 & 1.00 & 1 & 0.861 & 0.650 & 0.461 & 1.0 & 3.2 & 9.9 & 40 & 88 & 52 & 54 & 1.28 & 7.7 \\
\texttt{knn\_n25} & 25 & 21 & 84 & 10 & 1.00 & 1 & 0.697 & 0.292 & $6.4\mathrm{e}{-3}$ & 47.8 & 14.3 & 33.7 & 160 & 167 & 55 & 48 & 2.33 & 6.3 \\
\texttt{multiplier\_n15} & 15 & 13/14 & 222 & 7 & 1.03 & 3 & 0.452 & 0.066 & $6.9\mathrm{e}{-4}$ & 79.5 & 29.8 & 85.3 & 331 & 443 & 41 & 36 & 1.90 & 5.2 \\
\texttt{qaoa\_n10} & 10 & 7 & 24 & 0 & 1.00 & 1 & 0.930 & 0.804 & 0.506 & 4.5 & 2.4 & 9.0 & 27 & 47 & 19 & 18 & 1.65 & 6.1 \\
\texttt{qaoa\_n6} & 6 & 4 & 36 & 0 & 1.00 & 1 & 0.897 & 0.738 & 0.538 & 5.5 & 3.5 & 14.4 & 36 & 71 & 14 & 18 & 3.39 & 11.0 \\
\texttt{qft\_n10} & 10 & 8 & 90 & 7 & 1.00 & 1 & 0.707 & 0.494 & 0.192 & 46.2 & 6.6 & 32.9 & 69 & 179 & 26 & 28 & 3.40 & 10.5 \\
\texttt{qft\_n18} & 18 & 15 & 294 & 45 & 1.00 & 1 & 0.253 & 0.100 & $2.3\mathrm{e}{-3}$ & 208 & 19.1 & 99.4 & 195 & 587 & 38 & 55 & 1.95 & 5.9 \\
\texttt{qft\_n4} & 4 & 3 & 12 & 0 & 1.00 & 1 & 0.965 & 0.906 & 0.826 & 3.0 & 1.0 & 4.1 & 12 & 23 & 9 & 7 & 2.14 & 3.3 \\
\texttt{qpe\_n9} & 9 & 7 & 43 & 3 & 1.00 & 1 & 0.851 & 0.662 & 0.261 & 22.8 & 5.9 & 18.4 & 53 & 85 & 15 & 23 & 2.25 & 9.0 \\
\texttt{shor\_n5} & 5 & 3 & 25 & 0 & 1.00 & 1 & 0.928 & 0.757 & 0.600 & 6.2 & 4.4 & 10.5 & 46 & 49 & 7 & 11 & 0.94 & 8.0 \\
\texttt{wstate\_n27} & 27 & 23 & 52 & 0 & 1.00 & 1 & 0.855 & 0.461 & 0.072 & 7.0 & 10.3 & 18.4 & 103 & 103 & 43 & 53 & 0.71 & 6.8 \\
\midrule
Geo.\ mean & & & & & & & 0.746 & 0.357 & 0.020 & 11.8 & 6.0 & 16.0 & & & & & 1.45 & 6.6 \\
\bottomrule
\end{tabular}
\end{table*}

\subsection{Overall Performance}
\label{sec:eval_overall}

\textbf{Resource requirements.} BRIDGE-F uses $1.77\times$ as many atoms as the baselines in geometric mean, with per-circuit ratios of $1.57\times$--$1.87\times$ (Table~\ref{tab:resources}). Across the suite, it uses 657 atoms versus 365 for each baseline. The largest circuit, \texttt{adder\_n64}, uses $64+53=117$ atoms. 

ZAP and Enola use 659 and 840 trap sites across the suite, respectively, including additional sites for entanglement and partner atoms. Their spatial footprints are $1.45\times$ and $6.62\times$ BRIDGE-F's, respectively, in geometric mean. BRIDGE thus trades a larger atom population for a smaller spatial footprint on average.

\begin{figure*}[t]
  \centering
  \includegraphics[width=\linewidth]{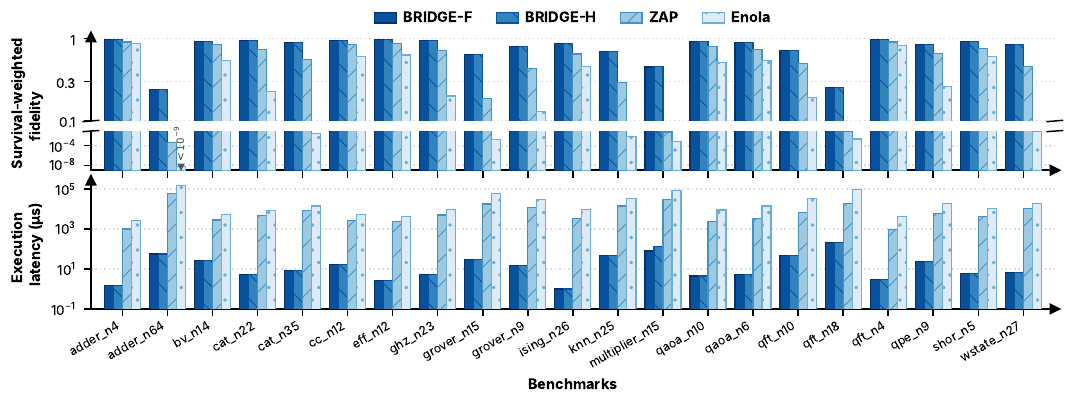}
  \caption{Overall co-design performance: $F_{\mathrm{sw}}$ (upper panels; higher is better) and execution latency (lower panel; lower is better). All values are model estimates excluding circuit single-qubit costs. The quality axis has two logarithmic segments that meet at $0.1$. Enola's \texttt{adder\_n64} quality, $1.3\times10^{-17}$, is shown as a marked stub below the $10^{-9}$ floor.}
  \vspace{-2ex}
  \label{fig:overall_performance}
\end{figure*}

\textbf{Performance.} At nominal costs, F and H achieved higher modeled quality than ZAP and Enola on all 21 circuits (Figure~\ref{fig:overall_performance}). F's geometric-mean quality ratios were $2.09\times$ relative to ZAP and $37.5\times$ relative to Enola, with medians of $1.30\times$ and $3.68\times$. The smallest quality ratio relative to ZAP was $1.07\times$, on \texttt{qft\_n4}. F reduced modeled execution latency by geometric-mean factors of $509$ and $1355$, respectively, with at least a $91$-fold reduction on every circuit.

The large advantage on \texttt{adder\_n64} substantially increased the geometric-mean quality ratios. In both transport-based designs, its 64 atoms remained exposed throughout 62--153~ms executions involving hundreds of transports. Excluding this circuit reduced F's quality ratios to $1.57\times$ relative to ZAP and $6.91\times$ relative to Enola. The primary comparison retains all 21 circuits, with medians indicating the typical gain.

Higher atom survival accounted for part of the modeled quality advantage. Geometric-mean survival probabilities were $0.9999$ for F, $0.882$ for ZAP, and $0.815$ for Enola. Survival therefore contributed quality-ratio factors of $1.13\times$ and $1.23\times$ relative to the respective baselines. The remaining advantage arose from conditional fidelity, including reduced handover error and idle decoherence. In Enola, Rydberg exposure of idle atoms contributed an additional geometric-mean quality factor of $0.09$.

QFT-18 illustrates why SWAP routing is necessary for distant interactions. Relay-only routing averaged 2.14 buffers per CZ, or 3.3 modules and 4.6 Hadamards, yielding quality $0.036$ compared with ZAP's $0.0997$. F inserted 45 SWAPs, reducing every relay to one buffer. The resulting 429 one-module relays and 270 SWAP Hadamards replaced 966 modules and 1344 Hadamards. Quality reached $0.253$, approximately $2.5\times$ ZAP's value, with execution latency of $208~\mu\mathrm{s}$ compared with $19$~ms. Across the suite, F inserted 78 SWAPs on seven circuits with long-range or changing interaction patterns; local circuits required none.

After SWAP routing, 18 of 21 circuits used exclusively one-buffer relays. The remaining three averaged no more than 1.38 buffers per relay. The evaluated advantage therefore depended predominantly on single-buffer BAM performance.

\subsection{Sensitivity to Relay Costs}
\label{sec:eval_sensitivity}

\begin{figure*}[t]
  \centering
  \includegraphics[width=\textwidth]{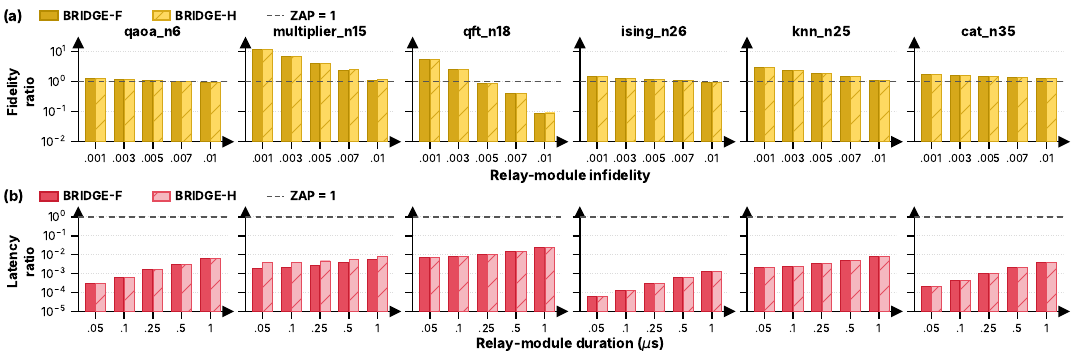}
  \caption{Sensitivity to relay costs across six circuits. (a)~Modeled quality ratio $F_{\mathrm{sw}}(\mathrm{F/H})/F_{\mathrm{sw}}(\mathrm{ZAP})$ versus module infidelity, with module duration fixed at $0.25~\mu\mathrm{s}$; higher is better. (b)~Modeled execution-latency ratio $T_{\mathrm{exec}}(\mathrm{F/H})/T_{\mathrm{exec}}(\mathrm{ZAP})$ versus module duration, with module infidelity fixed at $0.003$; lower is better. Each circuit uses a fixed ZAP reference at nominal parameters. Solid and hatched bars denote F and H, respectively; dashed lines mark parity. Vertical axes are logarithmic. Each bar represents one model estimate, excluding circuit single-qubit costs.}
  \vspace{-2ex}
  \label{fig:eval_relay_sensitivity}
\end{figure*}

We systematically varied module infidelity over $\{0.001,0.003,0.005,0.007,0.01\}$ and duration over $\{0.05,0.1,0.25,0.5,1\}~\mu\mathrm{s}$ on six circuits. These ranges extend to twice the baseline direct-CZ infidelity and four times the nominal module duration. Relay Hadamards retained their $1~\mu\mathrm{s}$ duration throughout the sweeps. Each point was compiled independently, with F and H sharing the same initial static plan. ZAP used a fixed nominal reference per circuit because its costs do not depend on relay parameters. Figure~\ref{fig:eval_relay_sensitivity} reports both sweeps for all six circuits. These analyses test sensitivity within the independent-error model, excluding correlated and coherent errors.

\textbf{Relay infidelity.} At nominal infidelity $0.003$, F/ZAP quality ratios were $6.89$ on multiplier, $2.54$ on QFT, $2.38$ on KNN, and $1.62$ on cat. Ratios for \texttt{ising\_n26} and \texttt{qaoa\_n6} were $1.32$ and $1.22$, respectively. At $0.007$, five circuits retained an advantage, with ratios of $2.45$ (multiplier), $1.51$ (KNN), $1.41$ (cat), $1.08$ (Ising), and $1.05$ (QAOA). QFT was most sensitive, falling below parity between $0.003$ and $0.005$ and reaching $0.39$ at $0.007$. At infidelity $0.001$, the ratios increased to $11.5$ on multiplier and $5.65$ on QFT. At $0.01$, cat ($1.27$), multiplier ($1.13$), and KNN ($1.07$) remained ahead, whereas Ising and QAOA fell to $0.93$--$0.94$. H accepted moves only on multiplier, improving quality by at most $3.5\%$; elsewhere, H and F were identical.

At infidelity $0.005$, relay modules and baseline direct CZs had equal assumed quality. Recompilation preserved F's advantage on five of six circuits ($1.13$--$4.12\times$), but QFT fell to $0.86\times$ ZAP (Figure~\ref{fig:eval_relay_sensitivity}a). By contrast, rescoring nominal plans at this infidelity preserved an advantage on all 21 circuits. The geometric-mean ratio was $1.73\times$, with a minimum of $1.04\times$ and $1.07\times$ on QFT-18. This distinction exposes a nonmonotonic response of F's SWAP heuristic to changing costs. The fixed-plan comparison nevertheless shows that the nominal advantage need not depend on superior relay-module quality.

\textbf{Relay duration.} At infidelity $0.003$, all six circuits retained at least a $41$-fold latency advantage throughout the duration sweep (Figure~\ref{fig:eval_relay_sensitivity}b). At $1~\mu\mathrm{s}$ per module, F's latency reached at most $2.4\%$ of ZAP's, corresponding to $459~\mu\mathrm{s}$ on QFT. Over the $20\times$ duration range, QFT latency increased by $3.2\times$, because relay and SWAP Hadamard durations remained fixed. H accepted two moves on multiplier at every duration and none elsewhere. These moves increased latency from F's $57$--$175~\mu\mathrm{s}$ to $117$--$241~\mu\mathrm{s}$ for a $0.8\%$ quality improvement. Quality changed by at most $0.4\%$ across the duration sweep, with the largest variation on QFT.

We also tested stricter serialization to assess sensitivity to concurrent-relay assumptions. The nominal scheduler permits relays whose atoms lie within one another's blockade range. Retiming exported F plans to separate relays with atoms within $10~\mu\mathrm{m}$ increased latency by $16\%$ in geometric mean, with a maximum threefold increase. F nevertheless retained a $437$-fold geometric-mean latency advantage over ZAP, or $342$-fold under fully serial execution. These scheduling checks quantify serialization costs without validating physical crosstalk suppression.

Joint parameter scans confirmed greater sensitivity to relay infidelity than duration. A coarse $5\times5$ scan of \texttt{knn\_n25} and \texttt{qft\_n18} identified a crossover only for QFT. We therefore refined QFT sampling to a $13\times9$ grid, covering infidelities $0.003$--$0.006$ in increments of $0.00025$ and nine durations from $0.05$ to $1~\mu\mathrm{s}$. All 117 F compilations completed, with at most $1\%$ quality variation across durations at each sampled infidelity. Increasing infidelity from $0.003$ to $0.006$ reduced quality fourfold at each duration. F exceeded ZAP at every duration through infidelity $0.0045$ ($1.21\times$) and remained within $1\%$ at $0.00475$ and $0.00525$. It fell below ZAP at $0.005$ ($0.86\times$) and from $0.0055$ onward. The nonmonotonic crossover near $0.005$ therefore reflects both primitive costs and heuristic plan selection.

This sensitivity follows from accumulated module error in the cost model. The factor $(1-\epsilon_p)^{N_p}$ compounds over $N_p=\sum_g(2L_g-1)$ modules, whereas longer duration adds idle decoherence and background exposure. The latter contributions remain small at the assumed $T_2=1.5$~s and $20$~s lifetimes.

\subsection{Atom Loss}
\label{sec:eval_loss}

We evaluate how atom loss affects survival during execution and how pre-run vacancies affect compilation.

\begin{figure}[t]
  \centering
  \includegraphics[width=\columnwidth]{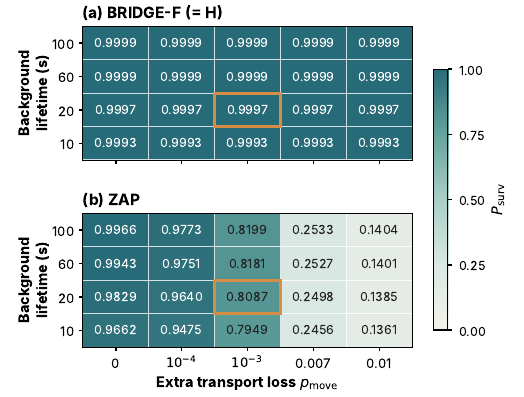}
  \caption{Atom survival on \texttt{qft\_n18}, with a shared $[0,1]$ color scale. Rows give the shared Rb/Cs background lifetime; columns give the additional loss probability per complete transport episode. Orange outlines mark the nominal parameters $(20~\mathrm{s},10^{-3})$. H accepts no move in any cell, so its plans are identical to F's. Each cell is a model estimate rounded to four decimals.}
  \vspace{-2ex}
  \label{fig:eval_loss}
\end{figure}

\textbf{Background and transport-induced loss.} We use \texttt{qft\_n18} to illustrate how repeated atom transport affects survival. We vary the shared background lifetime over $\{10,20,60,100\}$~s and the additional loss probability per transport over $\{0,10^{-4},10^{-3},0.007,0.01\}$ (Figure~\ref{fig:eval_loss}). To assess whether the trend extends across the suite, we evaluate all 21 circuits at five settings. These comprise transport-loss probabilities of $0$, $10^{-3}$, and $0.01$ at a 20~s lifetime, and lifetimes of 10 and 100~s at nominal transport loss. At each setting, we recompile F and H and rescore ZAP and Enola (Table~\ref{tab:loss_suite}).

H accepted no moves on QFT-18 at any sampled loss setting. BRIDGE therefore avoided transport loss, with survival probabilities of $0.9993$--$0.9999$ across the tested lifetimes. At nominal lifetime, ZAP's survival declined from $0.983$ without transport loss to $0.809$ at $p_{\mathrm{move}}=10^{-3}$ and $0.139$ at $0.01$. Its 195 transports contribute the survival factor $(1-p_{\mathrm{move}})^{N_{\mathrm{move}}}$. BRIDGE's quality remained approximately $0.253$, exceeding ZAP by $2.1\times$ without transport loss and $15\times$ at $0.01$.

\begin{table}[t]
\centering
\caption{Survival and quality under varied loss parameters across all 21 circuits. $P_{\mathrm{surv}}$: F's minimum and the baselines' geometric mean. Gain: F's $F_{\mathrm{sw}}$ relative to each baseline, reported as geometric mean (median). H equals F except on \texttt{multiplier\_n15}.}
\label{tab:loss_suite}
\footnotesize
\setlength{\tabcolsep}{4.2pt}
\begin{tabular}{@{}ll|rrr|rr@{}}
\toprule
 & & \multicolumn{3}{c|}{$P_{\mathrm{surv}}$} & \multicolumn{2}{c}{Gain of F over} \\
\multicolumn{2}{@{}l|}{Condition} & F (min) & ZAP & Enola & ZAP & Enola \\
\midrule
Transport & $0$ & 0.9997 & 0.984 & 0.959 & 1.88 (1.24) & 31.9 (3.08) \\
loss (20 s) & $10^{-3}$\,$^\ast$ & 0.9997 & 0.882 & 0.815 & 2.09 (1.30) & 37.5 (3.68) \\
 & $10^{-2}$ & 0.9997 & 0.328 & 0.187 & 5.63 (1.93) & 164 (6.87) \\
\midrule
Lifetime & 10 s & 0.9993 & 0.868 & 0.782 & 2.13 (1.30) & 39.1 (3.74) \\
(loss $10^{-3}$) & 100 s & 0.9999 & 0.893 & 0.843 & 2.07 (1.29) & 36.3 (3.63) \\
\bottomrule
\multicolumn{7}{@{}l}{$^\ast$\,Nominal.}
\end{tabular}
\vspace{-2ex}
\end{table}

F maintained survival of at least $0.9993$ and exceeded both baselines in quality on every circuit across all five loss settings. Its geometric-mean quality ratio relative to ZAP increased from $1.88\times$ without transport loss to $2.09\times$ at nominal loss and $5.63\times$ at $0.01$. At the highest transport loss, geometric-mean survival fell to $0.33$ for ZAP and $0.19$ for Enola. Relative to Enola, F's quality ratio increased from $31.9\times$ to $164\times$. Varying background lifetime from 10 to 100~s changed the quality ratios by less than $3\%$ relative to ZAP and $8\%$ relative to Enola.

\begin{figure}[t]
  \centering
  \includegraphics[width=\columnwidth]{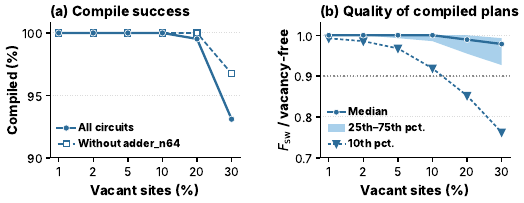}
  \caption{Compilation with pre-run vacancies. A random fraction of data and buffer sites is removed before compilation (21 circuits $\times$ 20 samples per rate). (a)~Fraction of samples that compile successfully. (b)~Quality relative to the vacancy-free plan among successful compilations: median, 25th--75th percentile band, and 10th percentile. The dotted line marks $0.9$.}
  \label{fig:eval_vacancies}
\end{figure}

\textbf{Pre-run vacancies.} To test adaptation to missing atoms before execution, we randomly remove 1--30\% of the candidate data and buffer sites before compilation. For each rate and circuit, we generate 20 samples and let F compile around the vacancies (Figure~\ref{fig:eval_vacancies}).

F compiled all 420 samples at each vacancy rate through 10\%, with unchanged median quality and latency. At 10\% vacancy (8 data and 10 buffer sites), 91\% of samples retained quality within 10\% of the vacancy-free value. At 20\% and 30\%, respectively, 418 and 391 samples compiled successfully. Among successful runs, median quality was $0.989$ and $0.979$ of the vacancy-free value. Of the 29 failures at 30\%, 16 involved \texttt{adder\_n64}, which requires 64 of the 81 data sites. Median recompilation time was $0.1$~s, with a maximum of $1.1$~s. These results establish adaptation before execution; they do not demonstrate recovery from atom loss during a shot.

\subsection{Ablation Studies}
\label{sec:eval_placement}

The comparison with ZAP and Enola changes hardware and compiler simultaneously. To isolate component contributions, we vary routing, mobility, relay construction, or placement while keeping the BRIDGE fabric and evaluator fixed. We use Graph-Greedy F as the reference and evaluate all 21 circuits at nominal costs (Table~\ref{tab:eval_compiler}).

\begin{table}[t]
  \centering
  \caption{Ablations of BRIDGE-F on the same fabric, 21 circuits. Entries are cross-circuit geometric-mean ratios to Graph-Greedy F (for the seeded searches, mean $\pm$ sample standard deviation over three seeds). $T_{\mathrm{comp}}$ includes training for ML; $Q=F_{\mathrm{sw}}$ and $T_{\mathrm{exec}}$ exclude circuit single-qubit costs. ZAP is shown for reference.}
  \label{tab:eval_compiler}
  \footnotesize
  \setlength{\tabcolsep}{4pt}
  \begin{tabular}{@{}llccc@{}}
    \toprule
    Part & Variant & $Q\uparrow$ & $T_{\mathrm{exec}}\downarrow$ & $T_{\mathrm{comp}}\downarrow$ \\
    \midrule
    -- & BRIDGE-F & $1$ & $1$ & $1$ \\
    Routing & Relay only & $0.875$ & $1.71$ & $0.60$ \\
     & SABRE & $0.971$ & $1.00$ & $1.90$ \\
    Mobility & BRIDGE-H & $1.000$ & $1.03$ & $7.0$ \\
    Relay & Without BAM & $0.516$ & $13.6$ & $0.93$ \\
    Placement & Greedy+CBO & $1.002{\pm}0.001$ & $0.955{\pm}0.015$ & $34.9{\pm}1.1$ \\
     & Greedy+ML+CBO & $1.002{\pm}0.001$ & $0.956{\pm}0.013$ & $35.7{\pm}1.3$ \\
    \midrule
    Reference & ZAP & $0.478$ & $509$ & -- \\
    \bottomrule
  \end{tabular}
\end{table}

\textbf{Routing.} We compared F's combined SWAP and relay routing with relay-only routing and SABRE~\cite{sabre2019}. SABRE places qubits and inserts SWAPs on the graph of data sites sharing a buffer. Its output uses F's scheduler, quality model, and SWAP-Hadamard costs. Removing SWAPs reduced geometric-mean quality by $12.5\%$ and increased execution latency to $1.7\times$ F's, primarily on circuits with distant interactions. On QFT-18, quality fell from $0.253$ to $0.036$. SABRE matched F on nine local circuits and underperformed on eight, with F's largest advantage of $1.65\times$ on \texttt{adder\_n64}. SABRE led by at most $2.5\%$ on four small circuits owing to better placement. These comparisons support combining SWAPs with relays for long-range interactions, while showing that F does not dominate every placement.

\textbf{Mobility.} Physical relocation provided limited additional benefit over F at nominal costs. H accepted two moves on \texttt{multiplier\_n15}, improving quality by $0.78\%$ but increasing latency from $79.5$ to $138.7~\mu\mathrm{s}$. No moves were accepted elsewhere after F's SWAP routing had shortened long relays. Starting H from relay-only plans also failed to match F. Across the seven circuits requiring SWAPs, H accepted 37 moves but achieved $1.35$-fold lower geometric-mean quality, with a $4.4$-fold deficit on QFT-18. Although movement incurs less local error than a SWAP, H considers individual scheduling events, whereas SWAP routing uses a 20-gate window and layout refinement. Initialized from F, H preserves or improves quality by construction, at $7.0\times$ the geometric-mean compilation time.

\textbf{Without BAM.} We replaced BAM relays with CNOT propagation through all $L$ buffers followed by a direct buffer--target CZ and uncomputation. This construction requires $2L+1$ modules and $4L$ Hadamards; SWAPs use the same replacement relays. Relative to BAM, each relay adds two modules and four Hadamards, increasing a one-buffer relay from one module to three. Under the stated costs, removing BAM approximately halved quality and increased latency $13.6$-fold. The replacement uses two-atom gate primitives and achieved $1.08\times$ ZAP's geometric-mean quality, leading on 13 of 21 circuits. Its geometric-mean execution latency remained lower by a factor of $37$.

\textbf{Placement.} We compared Graph-Greedy with Greedy+CBO and Greedy+ML+CBO, using three seeds per search variant. Following forward--backward routing refinement, population search yielded only small additional quality gains. Greedy+CBO preserved or improved quality on every circuit, with improvements on 11--13 of 21 circuits across seeds. The geometric-mean gain was $0.24\%$, reaching at most $2.2\%$ on multiplier and QFT-18. Execution latency decreased by $4.5\%$, while geometric-mean compilation time increased approximately 35-fold. Total compilation took approximately 8 minutes for the suite, compared with $3.4$~s for Graph-Greedy. The learned initializer changed outcomes by less than the variation across seeds. Training required less than one second per circuit and is included in $T_{\mathrm{comp}}$. These limited gains support training-free Graph-Greedy as the default for the tested circuits and costs.

\subsection{Scalability}
\label{sec:eval_scalability}

\begin{figure*}[t]
  \centering
  \includegraphics[width=\linewidth]{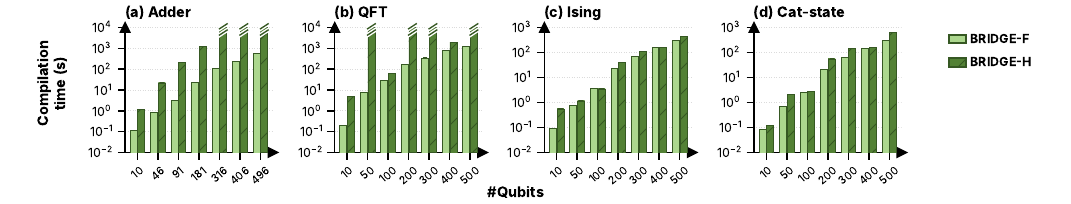}
  \caption{Compilation time across four circuit families. Each panel uses an independently scaled logarithmic time axis; horizontal positions are equally spaced and labeled with actual qubit counts. Light and dark bars denote F and H, respectively. Full-height bars with break marks indicate H runs that exceeded the 1-hour limit (adder-316/406/496, QFT-50/200/300/500); their heights do not represent measured times. All other bars show a single wall-clock measurement.}
  \vspace{-2ex}
  \label{fig:eval_scalability}
\end{figure*}

We evaluated Graph-Greedy F and H on 28 circuits spanning 10--500 qubits. Adder instances contained 10, 46, 91, 181, 316, 406, and 496 qubits. QFT, Ising, and cat-state instances each contained 10, 50, 100, 200, 300, 400, and 500 qubits. Above 81 qubits, data and buffer grids were expanded to near-square arrays with unchanged spacing and cost parameters. The 500-qubit instances used 506 data sites and 552 buffer sites. Each compilation was timed once with a 1-hour limit (Figure~\ref{fig:eval_scalability}).

F completed all 28 circuits within the specified time limit. The 500-qubit Ising and cat circuits required $313$ and $293$~s, respectively; the 496-qubit adder required $565$~s. The 500-qubit QFT required $1232$~s, inserting $5{,}767$ SWAPs for $14{,}261$ CZs. Mean relay length remained at most $1.22$ buffers across the tested sizes.

H completed 21 circuits and timed out on the 316-, 406-, and 496-qubit adders and QFT instances with 50, 200, 300, and 500 qubits. Width alone did not determine completion: QFT-100 and QFT-400 completed because no move candidates reached full evaluation. Candidate evaluation and rescheduling therefore constrained H's scalability in these tests. Among completed runs, one move on each of the 200- and 300-qubit cat circuits improved quality by $3.0\%$ and $9.0\%$, respectively. Because H starts from a complete F plan, that plan remains available if mobility search times out. Compilation at these scales does not establish useful physical execution fidelity.

\section{Related Work}
\label{sec:related}

Reconfigurable neutral-atom compilers transport atoms to bring interacting qubits within gate range. OLSQ-DPQA combines SMT-based layout synthesis with greedy heuristics for larger circuits~\cite{dpqa2024}, whereas Enola separates placement, routing, and edge-coloring-based scheduling~\cite{tan2025}. Atomique combines coarse- and fine-grained mapping with parallel-gate scheduling~\cite{wang2024}. DasAtom transports atoms between mappings of partitioned subcircuits~\cite{huang2025dasatom}, while Parallax schedules movement without SWAPs~\cite{parallax2024}. These methods satisfy coupling constraints by relocating atoms carrying quantum states.

Zoned architectures organize transport among storage, entangling, and measurement regions. ZAC reuses qubits to reduce inter-zone traffic~\cite{lin2025}, while PowerMove jointly optimizes allocation, scheduling, and inter-zone communication~\cite{powermove2024}. ZAP combines a zoned layout with a parallelizable compilation workflow~\cite{huang2026}. Related approaches address routing-aware placement~\cite{stade2025routing}, goal-directed search at thousand-qubit scales~\cite{stade2025search}, and physics-aware partitioning~\cite{pac2025}. Cross-compiler evaluation~\cite{khusainov2026} and multi-qubit gate synthesis~\cite{huang2026mobius} address complementary aspects of neutral-atom compilation. BRIDGE instead supplies connectivity through a fixed network of buffer atoms.

On fixed arrays, SWAP routing relocates quantum states while atoms remain stationary. General routers, including SABRE~\cite{sabre2019} and t$|$ket$\rangle$~\cite{tket2020}, insert SWAPs to satisfy coupling constraints. For neutral atoms, Baker et al.\ use long-range Rydberg interactions to shorten SWAP chains and tolerate atom loss~\cite{baker2021}. Geyser combines fixed-array mapping with multi-qubit gate synthesis to reduce SWAP overhead~\cite{geyser2022}, and Atomique evaluates fixed-array baselines alongside reconfigurable arrays~\cite{wang2024}. BRIDGE-F applies SWAP routing to a dual-species buffer fabric, using three one-buffer relays and six local Hadamards per SWAP.

Auxiliary atoms provide another mechanism for coupling stationary data qubits. Q-Pilot uses movable ``flying ancillas'' to mediate two-qubit gates~\cite{qpilot2024}. Cesa and Martin instead transfer a Rydberg excitation along an ancilla chain to entangle non-adjacent atoms~\cite{cesa2017}. BAM uses ORMD to mediate a logical gate through a buffer atom~\cite{sun2024}. The closely related directional-transport (DT) compiler propagates Rydberg excitation along a preconfigured ancilla chain to implement remote CZ gates~\cite{kong2026}. AOD motion establishes the ancilla channel before these interactions occur.

BRIDGE combines a shared network of stationary second-species buffers with cost-aware state relocation. Relay paths require neither ancilla transport nor atom motion for channel setup. The compiler compares relaying with buffer-mediated SWAPs and data-atom transport using modeled survival-weighted fidelity. This comparison accounts for gate errors, idle decoherence, and atom loss.

\section{Conclusion and Outlook}
\label{sec:future}

BRIDGE replaces repeated data-atom transport with connectivity through a fixed buffer network. Combining relays with buffer-mediated SWAPs improved modeled survival-weighted fidelity over ZAP and Enola across all 21 circuits. Modeled execution latency decreased by geometric-mean factors of approximately $500$ and $1{,}350$, respectively. Same-fabric comparisons supported the routing strategy, with F matching or exceeding SABRE on most circuits. These gains apply to the stated co-design model, which excludes circuit single-qubit costs.

Physical mobility provided limited additional quality at nominal costs while substantially increasing compilation time. Together with F's completion of all tested instances up to 500 qubits, these findings support F as the default and H as an optional refinement. The predicted advantage depends principally on relay fidelity, making experimental calibration essential to assessing hardware feasibility. Such calibration must establish compatible interactions and joint endpoint-gate and buffer-recovery performance.

\label{sec:ftqc}Fault-tolerant workloads may require a different balance between relaying and transport. The staggered Rb--Cs layout matches the geometry of a rotated surface code~\cite{fowler2012surface}, assigning Rb atoms to data qubits and Cs atoms to measurement qubits. Nearest-neighbor Rb--Cs gates could support syndrome extraction without relays or transport. This requires species-selective mid-circuit Cs readout and reset~\cite{singh2023}, together with scheduling that prevents simultaneous excitation of adjacent Cs atoms within blockade range. Readout could limit cycle time, precluding direct extrapolation of noisy intermediate-scale quantum (NISQ) latency gains to quantum error correction (QEC).

Quantum low-density parity-check (QLDPC) codes introduce nonlocal checks that can require longer relays and competing buffer resources. Large-scale parallel transport~\cite{bluvstein2023}, extending H's AOD batches, may become more useful under these conditions. Evaluating this trade-off requires a dedicated QEC study incorporating measurement, reset, leakage, and erasure~\cite{ma2023erasure}. The present NISQ evaluation does not establish logical error rates.

\section*{Acknowledgments}
This work was supported by the National Key Research and Development Program of China (Grant Nos. 2024YFB4504002 and 2025YFE0217200). 


\FloatBarrier
\ifhpcahasreferences
\bibliographystyle{IEEEtranS}
\bibliography{refs}
\fi


\end{document}